# Thermopower in the strongly overdoped region of single-layer $Bi_2Sr_2CuO_{6+\delta}$ superconductor

Z. Konstantinović[1], G. Le Bras[1], A. Forget[1], D. Colson[1], F. Jean[2,3], G. Collin[3], M. Ocio[1], C. Ayache[1]

[1]SPEC, CEA Saclay, 91191 Gif sur Yvette, France
[2]LEMHE, CNRS UMR 8647, Bât 415, Université Paris Sud, 91405 Orsay, France
[3]LLB, CEA-CNRS, CEA Saclay, 91191 Gif sur Yvette, France

The evolution of the thermoelectric power S(T) with doping, p, of single-layer $Bi_2Sr_2CuO_{6+\delta}$ ceramics in the strongly overdoped region is studied in detail. Analysis in term of drag and diffusion contributions indicates a departure of the diffusion from the T-linear metallic behavior. This effect is increased in the strongly overdoped range (p~0.2-0.28) and should reflect the proximity of some topological change.

While the underdoped region of the phase diagram of cuprates has been systematically investigated, especially in relation with the pseudogap phenomena [1], there is still a lack of experimental results in the strongly overdoped region. Currently considered as a Fermi liquid (FL), the overdoped region (OD) shows, however, properties near the suppression of superconductivity at a hole concentration of p~0.27 that are still poorly understood. We cite as examples, the observation of a maximum in doping variation of both the susceptibility and the electronic specific heat in $La_{2-x}Sr_xCuO_4$ (LSCO) compounds [2, 3] or the disappearance of the excitations responsible for the drag contributions in the thermoelectric power in $Bi_2Sr_{2-y}La_yCuO_{6+\delta}$ (Bi(La)-2201) single crystals [4].

Several physical properties show untypical behavior in the OD region, not only in their change with doping, but also in their temperature dependence. Susceptibility [5], electronic specific heat [3] and Knight shift [6] show a linear T-dependence instead of the T-independent behavior expected in the conventional FL picture. It should be also noted the unusual T-power dependence of the resistivity $\rho=\rho_0+\alpha T^n$, (n<2) [7] and the presence of a maximum in the Hall coefficient $R_H(T)$ [8], while a $T^2$ behavior of $\rho$ and a constant $R_H$ are expected in the FL frame.

Thermoelectric power, S, is one of the transport properties complementary to the resistivity and Hall effect. Although unsolved until now, the hole doping dependence of $S_{290K}$ (S at T=290 K), given by the universal Obertelli-Cooper-Tallon law (OCT) [9], is currently used as a measure of the hole concentration p. The unusual temperature dependence of S, still under discussion, can be understood on the basis of two different contributions: a drag term arising from the existing excitations in the system and a second one from the diffusion of carriers [4, 10]. This diffusion contribution gives rise to the robust negative values of S in the OD region while $R_H$ maintains a positive sign which is inconsistent with a simple FL picture [11]. In this region, currently reported thermopower varies linearly with temperature (FL behavior). However, the S(T) behavior of Bi(La)-2201 compounds shows a deviation from this T-linear behavior [12], which have been also displayed by recent theoretical studies [13, 14]. The evolution of thermopower in the strongly overdoped region, in the absence of systematic studies, can give insight in the evolution of many unusual normal state properties.

The present work is focused on the doping evolution of the thermoelectric power on $Bi_2Sr_2CuO_{6+\delta}$ (Bi-2201) in the strongly overdoped region. In the most overdoped samples, where only diffusion thermopower is present, an important deviation from simple FL picture is observed. To access to the diffusion part in the case of less overdoped samples, the non-zero drag contribution is subtracted from total thermopower [4, 10]. The doping evolution of the diffusion term shows an increasing departure from metallic behavior, indicating an accentuation of the underlying narrow-band properties. The absence of CuO chains in these compounds permits to associate the observed properties directly to the $CuO_2$ planes.

Polycrystalline $Bi_2Sr_2CuO_{6+\delta}$ samples were prepared by the classical solid reaction method [15]. The non-substituted Bi-2201 compound is intrinsically strongly overdoped (0.20<p<0.28). Thermogravimetric techniques were used to explore oxygen non-stoichiometry. Absolute oxygen content was determined by reduction with hydrogen, while the oxygen exchange was studied under annealing temperatures between 300 °C and 700 °C with varying oxygen partial pressure between $1atm<P_{O2}<10^{-4}atm$ [15, 16]. Hole concentration can be adjusted through oxygen excess δ in a reversible way. All samples have the same cationic composition which allows to follow the change of the physical properties as a function of oxygen excess in the compound. The phase purity of the samples is controlled by X-ray diffraction. The critical temperature is determined from the onset of dc magnetization measurements. Detailed susceptibility measurements on samples with the same compositions will be published elsewhere [17]. The thermoelectric measurements were performed by a conventional steady flow technique. Temperature and voltage gradients were simultaneously measured by T-type thermocouples. Samples dimensions are typically 3x7x1 $mm^3$. The absolute error does not exceed 1 μV/K in the whole temperature range. The low $T_c$ values allow to investigate normal state properties over a large interval of T below room temperature.

Typical thermoelectric power $S(T)$ of $Bi_2Sr_2CuO_{6+\delta}$ in the overdoped region of their phase diagram $(T,\delta)$ is represented in Figure 1. The most overdoped sample with $T_c<1.5$ K is situated at the limit of the superconducting region ($p\sim0.28$), while the less overdoped sample is close to optimal doping ($p\sim0.20$). Excess oxygen values $\delta$ shown in the figure were experimentally obtained as described above, except for the two less overdoped samples where $\delta$ is determined by linear extrapolation of $S_{290}$ (see below) [15, 18]. The results are in good qualitative agreement with those obtained for Bi(La)-2201 ceramics [12], where the hole doping is modified by changing the cationic composition of the sample in the strongly overdoped region. The variation of $S(T)$ shows a complicated behavior, strongly dependent of oxygen excess $\delta$. One clear oscillation at low temperature is present in two states ($\delta=0.14$ and $\delta=0.13$), suggesting that S arises from two different contributions: $S=S_{drag}+S_{diff}$, where $S_{drag}$ is a positive drag term and $S_{diff}$ is a negative diffusion part. In the case of the most overdoped samples ($\delta=0.175$ and 0.18), $S(T)$ shows a marked curvature.

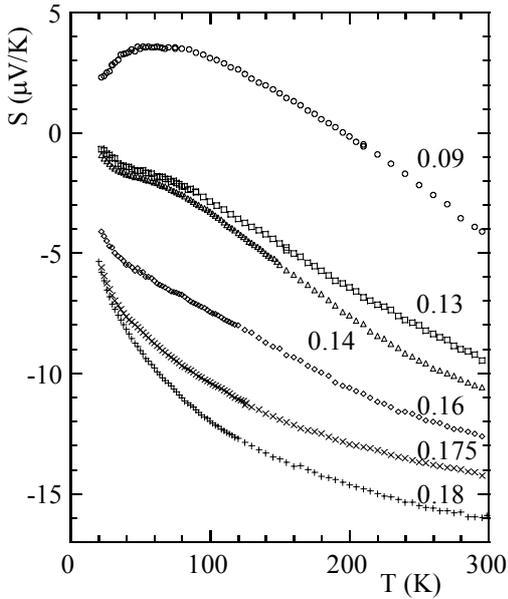

FIG. 1. Temperature dependence of the thermopower $S(T)$ for Bi-2201 polycrystalline samples in the overdoped region for indicated values of oxygen excess $\delta$.

The variation of the critical temperature as a function of room temperature thermopower, $S_{290K}$, is plotted in Fig. 2 for all measured samples. For comparison, the universal OCT relation: $S_{290K}=24.2-139p$ [19] included in the well-known phenomenological parabolic law: $T_c/T_{cmax}=1-82.6(p-0.16)^2$ [20] is represented as a solid line. The maximal critical temperature is estimated to be $T_{cmax}=20$ K which is a reasonable value for this one-layer system. Our data are slightly shifted ($\Delta p\sim0.025$, dotted line) from the universal behavior, as it was also reported in the case of Bi(La)-2201 single crystals [21]. All points $T_c$ vs $S_{290}$ can be reasonably fitted with a shifted parabolic line (dotted line), except for the least overdoped sample ($p\sim0.20$), which has lower $T_c$ value [22]. Consequently, in the absence of a more accurate determination, an estimation of the hole number p can be made from the OCT law [19] within an estimated error of 10% (values indicated at the top axis). In the inset of the same figure, we show $S_{290}$ values as a function of $\delta$ (circles). A linear fit, $S_{290}=6.82-122\delta$ (dashed line), assuming a linear relation between the hole number p and oxygen excess $\delta$, is used to extrapolate $\delta$ values for the two less overdoped samples (pluses).

In the case of $\delta=0.175$ and 0.18, the oscillation in $S(T)$ is absent indicating a zero drag term. Thus, only diffusion contributes to the total thermopower $S(T)=S_{diff}(T)$. This observed diffusion thermopower $S_{diff}(T)$ shows a deviation from the T-linear FL behavior. It can be described approximately by a phenomenological law of the form $T/S_{diff}=A+BT$ with two different parameters A, B<0. This behavior is illustrated in Fig. 3(a), where we have plotted $T/S_{diff}$ as a function of temperature for $\delta=0.175$ and 0.18.

With the goal to study the evolution of the diffusion thermopower, we must first analyze and substrate the drag contribution for the doping states with $\delta<0.175$. At high temperature (T>100 K), $S_{drag}$ is expected to be nearly constant [4,10] with a saturation value $S^0_{drag}$. Thus, we can assume that all of the T-dependence of $S(T)$ in this regime arises from $S_{diff}$. We can then extrapolate $S_{diff}(T)$ at low temperature and therefore deduce the shape of the low-temperature part of $S_{drag}$ which shows the characteristic step-like form [4,10]. An example of the above decomposition is shown for one doping state, $\delta=0.13$ in Fig. 3(b). Thus, this decomposition can naturally explain the observed dip at low T in $S(T)$ without involving an extra small peak in narrow band density of states as proposed in reference [13].

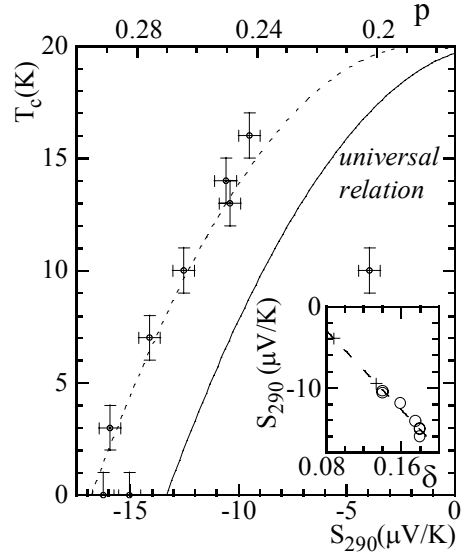

FIG. 2. $T_c$ vs thermopower at T=290 K, $S_{290}$. The solid line represents a universal behavior whereas the dotted line is a shifted parabolic law. In the insert, doping dependence of $S_{290}$. Circles correspond to samples with experimental $\delta$ values, whereas for pluses $\delta$ was determined by using a linear fit (dashed line). See main text for details.

The saturation value, $S^0_{drag}$, is directly proportional to the number of excitation modes per $CuO_2$, $n_{exc}$, responsible for this drag effect [4]. A comparison between our experimental data points and those obtained in reference [4] on Bi(La)-2201 single crystals (solid line) is done in Fig. 3(c). $T_c$ values are also indicated in the same figure. The number of dragging excitations decreases with increasing hole doping and vanishes close to p~0.28. Though the origin of this drag excitation is not clear until now, taking into account its doping variation, it is most probable that these excitations have a magnetic rather than a phonon origin.

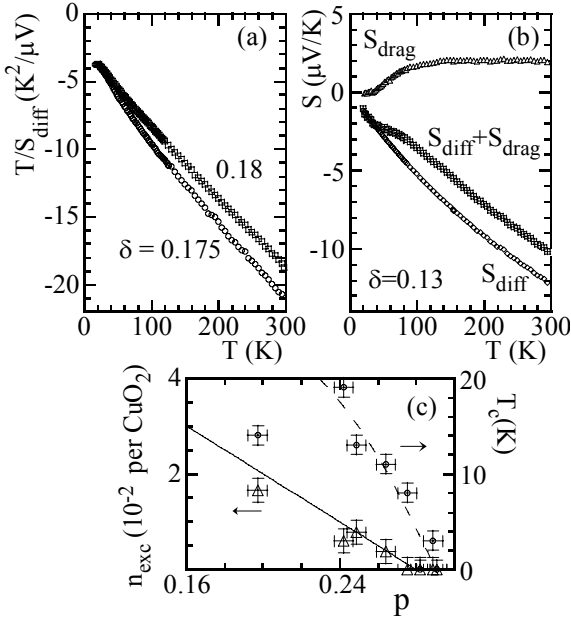

FIG. 3. (a) $T/S_{diff}$ vs T for $\delta=0.175$ and 0.18. (b) Extraction of the two different contributions of S(T), $S_{diff}$ and $S_{drag}$, for one typical state with $\delta=0.13$. (c) The number of excitations $n_{exc}$ and $T_c$ vs $\delta$. The results are compared with the previously determined variation [4] (solid line). The dashed line is a parabolic function (as mentioned in text).

Finally, once the above analysis is made, we can follow the evolution of the diffusion thermopower $S_{diff}$ with doping. In Fig. 4(a), we show the obtained linear T-dependence of $T/S_{diff}$ for T>100 K ($S_{diff}=T/(A+BT)$, see above). The parameter A is directly related to the slope $\alpha$ of the thermopower in the low temperature limit (T→0): $\alpha=A^{-1}$ (open circles in Fig. 4(b)). By using the Mott formula [23], $\alpha$ is also related to the slope of the conductivity, $\sigma$, at the Fermi level, $\alpha \sim 1/\sigma(\partial\sigma(\varepsilon_f)/\partial\varepsilon_f)$ where the principal contribution comes from the slope of the density of states $\partial\nu(\varepsilon_f)/\partial\varepsilon_f$. Its absolute value increases by more than a factor of 10 between the two doping limits in an exponential way ($0.035+1.8\times10^{-7}\exp(T/0.02)$, dotted line), indicating an accentuation of the narrow-band properties near the Fermi level. At the same time, B is a parameter characterizing the departure from linearity (open circles in Fig. 4(c)). In the case of the less overdoped state ($\delta=0.09$), $S_{diff}$ is nearly linear with T, i.e. B~0. With increasing doping, the parameter B increases by two orders-of-magnitude announcing the growing curvature in $S_{diff}(T)$. For comparison, the above analysis has also been extended to Bi(La)-2201 ceramics [12]. Corresponding results are plotted in Figure 4(b) and 4(c) as solid lines. As the diffusion thermopower is very sensitive to the electronic structure, the evolution of $S_{diff}$ with doping clearly indicates a general evolution in the electronic states with doping in these one-layer materials.

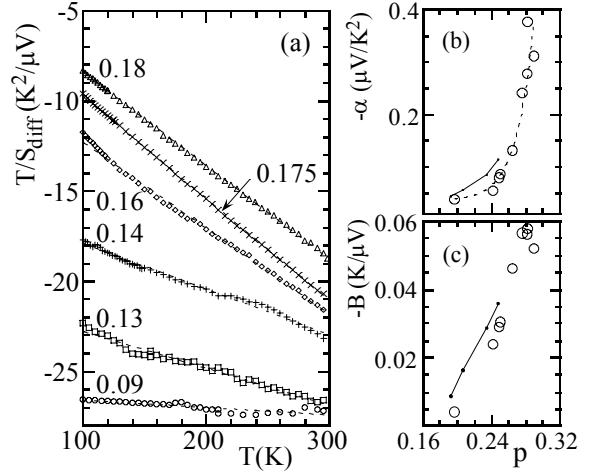

FIG. 4. (a). Evolution of the diffusion part of S(T) with doping for T>100 K (dashed lines are linear fits). (b) and (c) Respectively, evolution of the parameters $\alpha$ and B (open circles) with hole concentration. Analysis from results on Bi(La)-2201 ceramics [12] are also shown (close circles).

Any attempt to explain the above results should take into consideration a large hole-like Fermi surface in the strongly overdoped region in Bi(La)-2201 [24, 25] and $Bi_{1.8}Pb_{0.38}Sr_2CuO_{6-\delta}$ (Bi(Pb)-2201) [26] compounds and the existence of a saddle point, located at $(0,\pi)$ in the Brillouin zone [27]. Its position becomes closer to the Fermi surface with increasing doping [24] and it is located around 5 meV below the Fermi surface (FS) in the strongly overdoped case of Bi(Pb)-2201 ($T_c<4$ K) [26]. Thus, the saddle point, giving rise to a van Hove singularity (VHS) in the density of states (DOS) near the Fermi level, emphasizes narrow band properties.

A previous theoretical attempt to evaluate thermopower S(T) in the vicinity of a saddle point giving a non-linear T behavior in the OD region was reported in reference [28]. However, the zero thermopower value, S=0, was found when the Fermi level lies on the saddle point. In this case, the Fermi surface becomes electron-like on the OD side, which is not in agreement with photoemission measurements [24-26]. The negative electron-like diffusion thermopower and positive hole-like Hall coefficient in OD region can be explained within a tight-binding band model of the $CuO_2$ square lattice involving VHS [7]. The two-faced electron character in the perpendicular and parallel directions to the Fermi surface is in good agreement with the observed hole-like FS in the strongly OD region [24-26]. However, the Sommerfeld expansion used in reference [7] gives rise to a linear-T variation of $S_{diff}(T)$. The more accurate procedure, taking into account a rapid variation of the DOS in a few $k_BT$ (Fermi window) around Fermi level, leads to the deviation of the thermopower from this metallic behavior [13]. The obtained curvature of S(T) (when VHS lies around 5 meV below Fermi level) is less pronounced and has a smaller slope $\alpha \sim 0.1$ µV/K than in our case for the most overdoped sample ($\delta=0.18$). One recent theoretical work takes into account the different contributions to the thermopower along the MY (($0,\pi$)→($\pi,\pi$)) direction where band dispersion is very narrow with a low Fermi velocity $v_f$ (hot states) and along the ΓY (($0,0$)→($\pi,\pi$)) direction with wide band dispersion and high $v_f$ (cold states) [13]. The slope α, principally controlled by the density of states in the cold region, is still smaller ($\alpha \sim 0.08$ µV/K) than our determined values for strongly OD states. The evaluated S(T) in the OD region shows also deviations from T-linear behavior, but even less pronounced than in the previous model.

All these models include the narrow band properties and the existence of a peak in the density of states near the Fermi surface. This can be also indirectly probed from the susceptibility and the electronic specific heat measurements [2,3]. The increasing of susceptibility with doping, observed on samples with the same composition [17], indicates an enhancement of the DOS value at the Fermi level. This again emphasizes the growing narrow band properties, in good agreement with the increasing slope of the DOS seen in Figure 4(b). In the absence of a detailed evolution of band dispersions in the OD region, our measurements suggest a rapid approaching of the Fermi level to the van Hove singularity with doping and the expected topological change from hole-like to electron-like Fermi surface for slightly higher hole number (p>0.28). We should note, however, that the possible difference between experimental results and the above theoretical calculations could also arise from the incommensurate superstructure in the BiO layer and the related umklapp scattering [29].

In summary, we have presented the doping evolution of the thermoelectric power S(T) of Bi-2201 ceramics in the overdoped region. The diffusion thermopower $S_{diff}(T)$ shows a deviation from currently reported T-linear variation. The drag contribution to the total thermopower, as previously found, diminishes with increasing doping, suggesting its magnetic origin. After subtraction of this drag term, the evolution of the diffusion part from nearly T-linear behavior to a pronounced curvature indicates growing narrow band properties near the Fermi level and the proximity of a topological change of the Fermi surface.

We are grateful to C. Chaleil and L. Le Pape for their technical support. We thank M. Norman, C. Pepin, J. Bouvier, J. Bok, M. Roger and J.P. Carton for stimulating discussions.